\pdfoutput=1
\PassOptionsToPackage{table}{xcolor}
\documentclass[11pt]{article}

\usepackage[]{ACL2023}

\usepackage{times}
\usepackage{latexsym}
\usepackage{graphicx}
\usepackage{algorithm}
\usepackage{algorithmic}
\usepackage{multirow}
\usepackage{booktabs} 
\usepackage{amsmath} 
\usepackage[table]{xcolor}
\usepackage{comment}

\usepackage[T1]{fontenc}

\usepackage[utf8]{inputenc}

\usepackage{microtype}

\usepackage{inconsolata}

\title{ChatCite: LLM Agent with Human Workflow Guidance for Comparative Literature Summary}

\author{
    Yutong Li$^{1}$, Lu Chen$^{2,3}$, Aiwei Liu$^1$, Kai Yu$^{2,3}$,\textbf{Lijie Wen}$^{1}$\\
    \vspace{-0.6 ex} \small $^1$Tsinghua University, Beijing, China\\ \vspace{-0.9 ex}  \small $^2$X-LANCE Lab, Department of Computer Science and Engineering\\ \vspace{-0.9 ex} \small MoE Key Lab of Artificial Intelligence, SJTU AI Institute\\  \vspace{-0.6 ex}\small Shanghai Jiao Tong University, Shanghai, China\\ \small $^3$Suzhou Laboratory, Suzhou, China\\
 \small \texttt{li-yt21@mails.tsinghua.edu.cn}, \texttt{chenlusz@sjtu.edu.cn},  \texttt{wenlj@tsinghua.edu.cn}
  }

\begin{document}
\maketitle
\begin{abstract}
The literature review is an indispensable step in the research process. It provides the benefit of comprehending the research problem and understanding the current research situation while conducting a comparative analysis of prior works. However, literature summary is challenging and time consuming. The previous LLM-based studies on literature review mainly focused on the complete process, including literature retrieval, screening, and summarization. However, for the summarization step, simple CoT method often lacks the ability to provide extensive comparative summary. In this work, we firstly focus on the independent literature summarization step and introduce \emph{ChatCite}\footnote{Our code will be released after the review process.}, an LLM agent with human workflow guidance for comparative literature summary. This agent, by mimicking the human workflow, first extracts key elements from relevant literature and then generates summaries using a Reflective Incremental Mechanism. In order to better evaluate the quality of the generated summaries, we devised a LLM-based automatic evaluation metric, G-Score, in refer to the human evaluation criteria. The \emph{ChatCite} agent outperformed other models in various dimensions in the experiments. The literature summaries generated by \emph{ChatCite} can also be directly used for drafting literature reviews.
\end{abstract}

\section{Introduction}
As the rapid advancement of academic research, scholars must delve into existing literature to understand past studies, recognize future research trends, and find innovative approaches in their fields. Crafting a literature review entails searching for relevant literature and conducting detailed comparative summarization. It typically involves two main steps: literature collection followed by literature summary generation based on the collected sources. However, organizing a high-quality literature review necessitates scholars to engage in thorough analysis, organization, comparison, and integration of an extensive of related works, which is often a challenging and time-consuming task.

\begin{figure}
    \centering
    \includegraphics[width=1.0\linewidth]{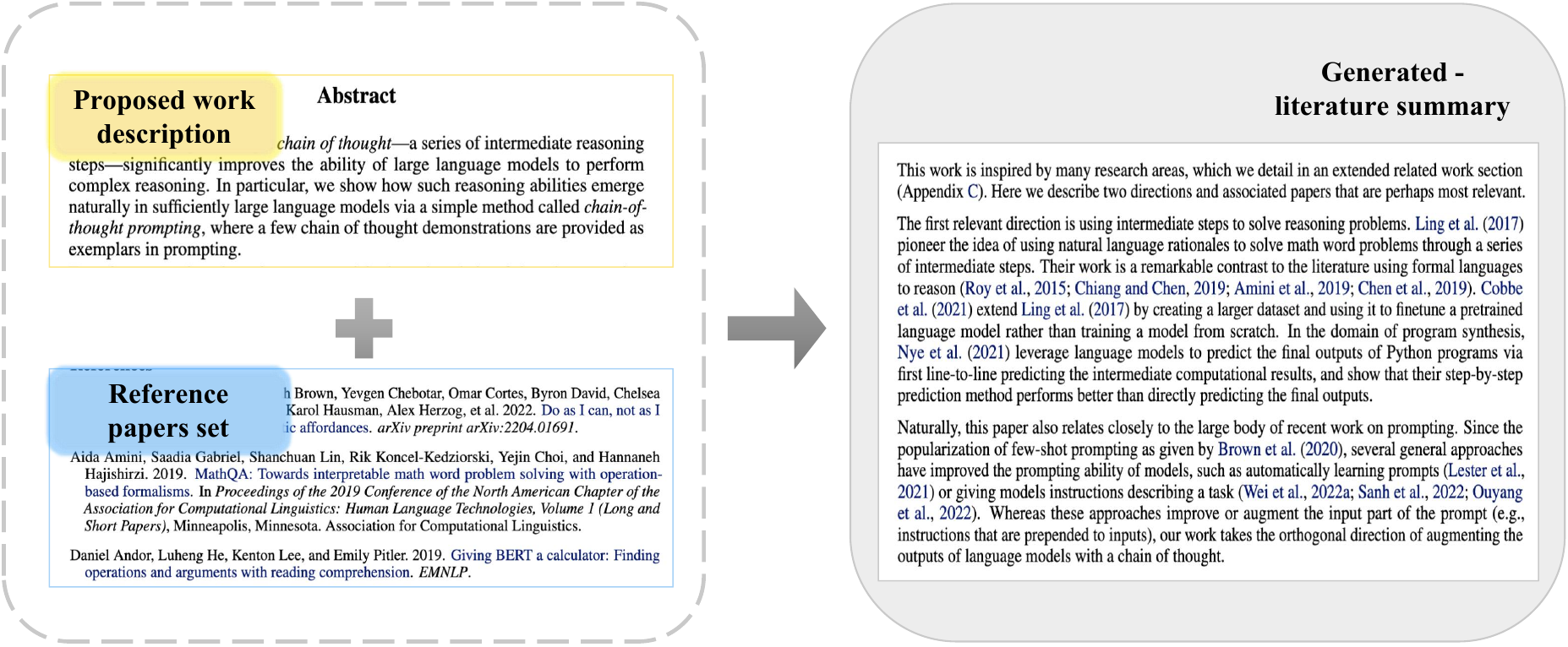}
    \caption{Literature Summary Task Description}
    \label{fig:task}
\end{figure}
Therefore, \citet{hoang-kan-2010-towards} have proposed the automatic generation of literature summary. However, machine-generated literature summaries often encounter challenges like information omission, lack of linguistic fluency, and insufficient comparative analysis. In traditional models, summaries generated through extraction and abstraction approach may miss key information due to the limitations of the model, leading to the lack of crucial points or findings of the generated summaries. Some automated systems may lack the ability for in-depth comparative analysis, potentially resulting in literature summaries that lack a comprehensive understanding of the relevant research in the field.

In recent years, with the rapid development of  large language models (LLMs) \cite{Radford2019LanguageMA, brown2020language}, their powerful capabilities in natural language generation tasks have been demonstrated across various tasks, that provides possibilities for handling longer texts and generating comprehensive summaries. Researchers have started exploring how to leverage LLMs to generate automatic literature summaries. \citet{wei2023chainofthought} propose a Chain-of-Thought (CoT) prompting method to enhance the ability of large language models to perform complex reasoning. CoT allows LLMs to devise their own plan, resulting in generated text that aligns more closely with human preferences.Recent study by \cite{huang2023role} and \citet{agarwal2024litllm} on literature review has focused more on how to retrieve relevant papers more accurately and neglected research on literature summarization. They use only simple CoT guidance to generate literature summaries, resulting in a lack of comparative and organizational analysis. Large language models, despite their fluent language generation, struggle to consistently produce comparative literature summaries due to their unpredictable an stochastic nature. The length limitations of these models require a two-step summarization approach, increasing the risk of information omission during abstract generation.

In this work, we focus on the independent literature summarization task, aiming to generate a comprehensive comparative literature summary through a certain collection of literature and a description of the proposed work, as illustrated in Figure \ref{fig:task}. 
To address these challenges mentioned above, our work proposes \emph{ChatCite}, a LLM-based agent guided by human workflow. Different from simple CoT prompting approach, the agent is designed with the human workflow guidance, rather than formulating the generation process in a black-box manner, ensuring a more stable generation of higher-quality generic summaries.

Furthermore, quality assessment for generative tasks has always been a challenge. Prior studies on literature summarization have primarily relied on text summarization metrics, such as ROUGE (\citet{Lin2004ROUGEAP}). However, traditional text summary evaluation metrics, like ROUGE, are not sufficient to assess the quality of literature summaries. More comprehensive evaluation criteria covering multiple dimensions are required to ensure that the generated literature summaries truly meet the requirements. Therefore, we combine human studies on literature reviews \cite{justitia2022automatic} to formulate the evaluation criteria for literature summaries from multiple dimensions \footnote{Six evaluation dimensions are: Consistency, Coherence, Comparative, Integrity, Fluency, Cite Accuracy.}, and propose an LLM-based automatic evaluation metric, G-Score. Experimental results demonstrate its consistency with human evaluations.

In this paper, we summarize our main contributions of our framework as follows:
\begin{itemize}
    \item we focus on the independent literature summarization step of literature review, and introduce \emph{ChatCite}, an LLM agent with human workflow guidance for comparative literature summary. 
    \item Based on research on literature summaries, we have developed a multidimensional quality assessment criterion for literature summaries. Additionally, we propose an LLM-based automatic evaluation metric, G-Score, demonstrating results consistent with human preferences.
    \item The experimental results indicate that \emph{ChatCite} outperforms other LLM-based literature summarization methods in all quality dimensions. The literature summaries produced by \emph{ChatCite} can be directly utilized for drafting literature reviews.
    \item We demonstrate that LLMs with human workflow guidance, have the ability to effectively perform comprehensive comparative summarization of multiple documents. Therefore, we infer that Large Language Models (LLMs) have the potential to handle more complex inferential summarization tasks.

\end{itemize}

\section{Related Work\footnote{Our related work utilizes summaries generated by ChatCite with GPT-4 as a draft, with minimal refinement. The information is comprehensive with minimal errors. The generated results organize the literature and include comparative analysis. The generated results are presented in the appendix (Table \ref{table:this_rw}).}}
In recent years, there is abundant research on generated literature summaries with the initial proposal made by \citet{hoang-kan-2010-towards}, to automate related work summarization created by a topic-related work summary based on an extractive approach. To generate citation sentence, \citet{Xing2020AutomaticGO} adopted a multi-source pointer-generator network with cross-attention mechanism, while \citet{10.1007/s11192-020-03630-2} utilized the ARWG system, employing a neural sequence learning process and \citet{ge-etal-2021-baco} proposed a BACO framework based on background knowledge and content. Furthermore, \citet{chen-etal-2021-capturing} employed the Relation-aware Related Work Generator (RRG) to generate citation paragraphs while \citet{Chen2022TargetawareAR} applied contrastive learning to generate target-aware related work segments. Yet traditional generation methods cannot generate the conprehensive coherent literature review due to the size of their model and the lack of the coherent and procedural language continuity.

Large Language Models (LLMs), such as GPT (\citet{Radford2019LanguageMA}, \citet{brown2020language}), have demonstrated their powerful capabilities in natural language generation tasks. The study by \citet{huang2023role} on the use of AI tools like ChatGPT in writing scientific review articles reveals the potential benefits and drawbacks of artificial intelligence in academic writing. Building on these insights, \citet{agarwal2024litllm} introduces the LitLLM toolkit, which overcomes challenges such as generating hallucinated content and overlooking recent research by adopting Retrieval Augmented Generation (RAG) principles, specialized prompting, and instructive techniques. However, these studies only applied a simple Chain of Thought (CoT) to the search and filtering process in literature reviews, resulting in poor readability. By comparison, \emph{ChatCite} focuses on the independent task of text summarization, aiming to generate higher-quality summaries. 

Furthermore, this paper introduced a multi-dimensional G-Score evaluation metric inspired by the previous attempt to use Large Language Models (LLMs) through chain-of-thought methods to evaluate the quality of natural language generation (NLG) systems (\citet{liu2023geval}, \citet{goyal2023news}) which is more consistent with human evaluation compared to traditional ROUGE metrics (\citet{lin-2004-rouge}).

\section{ChatCite}
\begin{figure*}[h]
  \centering
  \includegraphics[width=\textwidth]{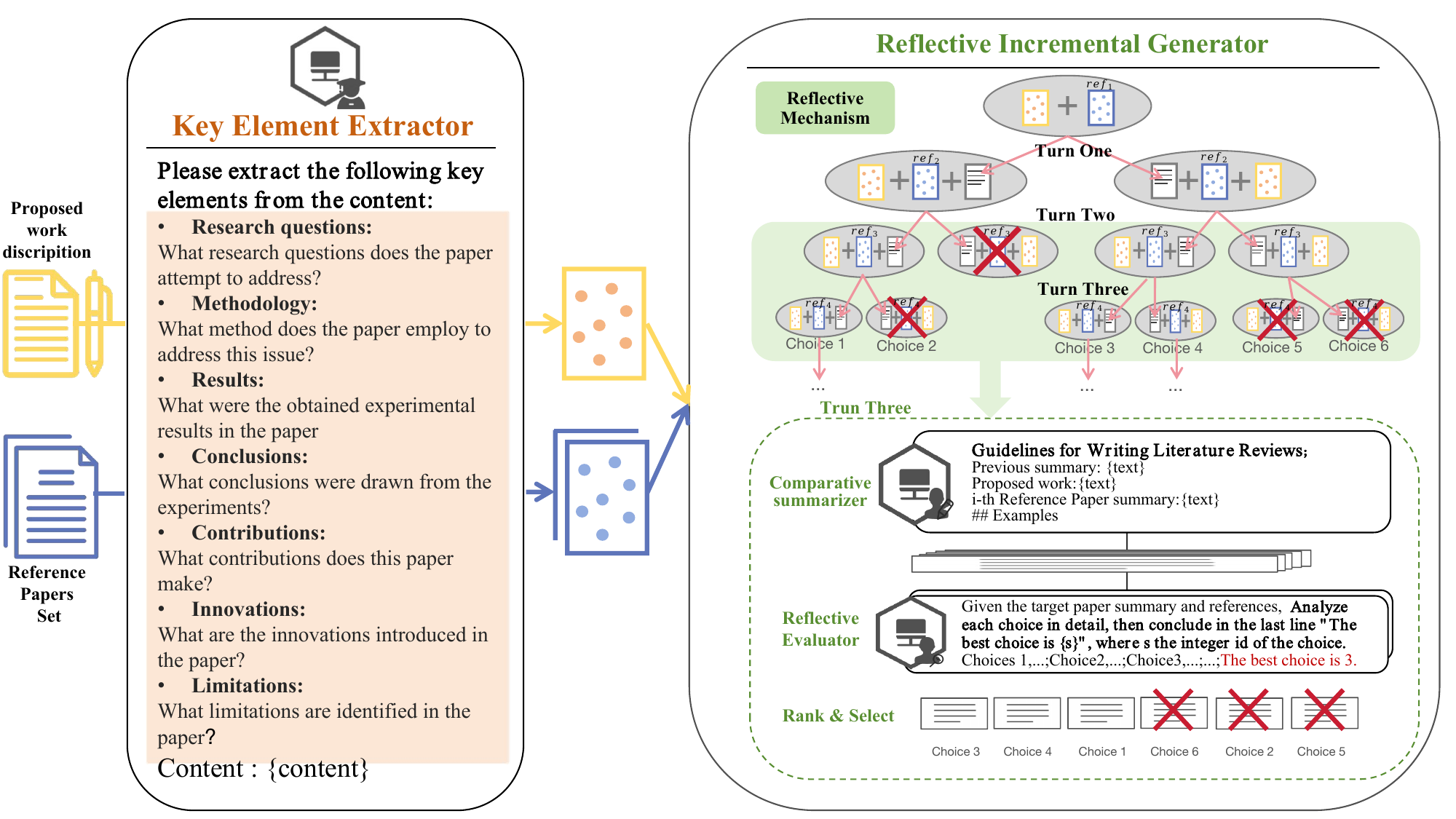}
  \caption{The \emph{ChatCite} consists of two modules, the Key Element Extractor and the Reflective Incremental Generator. The agent mimicking human workflow generates literature summary utilizing the Key Element Extractor to process the proposed work description and reference paper in Reference Papers Set. It then iteratively generates literature summaries using each paper in the Reference Papers Set, proposed work key elements and previous summary generated with the Reflective Incremental Generator.  This process is iteratively repeated until a complete related work summary is generated, and the optimal one is selected as the final result.}
  \label{fig:2}
\end{figure*}
The literature review task can be decomposed into two sub tasks: relevant papers retrieval and literature summaries generation. This work focuses on the independent task of literature summary generation.
Our task is to generate the literature summary based on the proposed work description $\textit{D}$ and a certain reference papers set $\textit{R}=\left\{r_1,r_2,...,r_n\right\}$. Given $\textit{D}$ and $\textit{R}$, our agent generates a literature summary $Y=f(\textit{D}, $\textit{R}$)$. 

Diverging from other types of summaries, such as news summaries, the literature summary generated directly by large language models using simple Chain-of-Thought (CoT) guidance in existing work mainly faces the following issues:

\noindent\textbf{Key Elements missing:} Because of the window limitations of LLMs, generating the complete literature review directly is challenging. Typically, a two-step approach is used involving summarization and literature review generation. However, this process can lead to the loss of key elements during summarization. Even if the entire literature summary can be directly generated, using the entire text may result in mistakes in understanding key elements and the loss of such elements.

\noindent\textbf{Lack of Comparative Analysis:} Comparative analysis is crucial in literature summary, requiring an analysis on the limitations and advantages of existing research methods, and focusing on differences and similarities in methods, experimental design, dataset usage, and more. Directly using CoT-generated results often lacks comparative analysis.

\noindent\textbf{Lack of Organizational Structure:} The literature summary generated solely by CoT tends to be discrete for each paper, lacking classification for similar works and an organized structure for the literature review.

To address these challenges, we have proposed an LLM agent for comparative literature summary with human workflow guidance, \emph{ChatCite}, consisting two modules: the Key Element Extractor and the Reflective Incremental Generator, as illustrated in Figure \ref{fig:2}. In this process, we utilize large language models as both generation and evaluation components, eliminating the need for additional model training and improving the quality of generated text to some extent. 

The generation process guided by human workflow is as follows: \begin{enumerate} 

\item The proposed work description and reference papers in the reference papers set are initially processed using the Key Element Extractor separately.

\item Iteratively generate literature summaries using reference papers set. In each iteration, use the comparative summarizer to generate a comparative analysis summary. Then, use the reflective evaluator to vote on the generated candidate results, ranking the vote score and retaining the top $n_c$ results. Iterate continuously until all reference papers are processed.
\end{enumerate}
The final output is selected based on the highest voting score among the generated related work summaries. 
    
In this section, we first elaborate on the specifics of the Key Element Extractor \textbf{(\S \ref{sec:2.2})} and the Reflective Iterative Generator module \textbf{(\S \ref{sec:2.3})} in detail.

\subsection{Key Element Extractor}
\label{sec:2.2}
 In order to retain sufficient key element for  literature summary, we create seven simple guiding questions based on  analysis \cite{justitia2022automatic} on literature review. We concatenate theses questions and the content required extraction as prompt to instruct LLMs extract the key elements. For each element, a simple question (shown in Figure \ref{fig:2}) is set to guide the model in extraction, and these questions are $Q_e=\left[q_1,q_2,...,q_7\right].$ These questions $Q_e$ and paper content $C$ are concatenated to form the key element extraction prompt 
 $P_e=\left[Q_e,C\right]$ . Using LLM as extraction decoder to extract key elements and storing them in memory.
 
\subsection{Reflective incremental Generator}
\label{sec:2.3}

To overcome the challenges of lacking comparative analysis and organizational structure in literature reviews generated by LLMs, we designed the reflective incremental generator. The generator uses the Comparative Summarizer to continue writing comparative summaries, combining the results from the previous turn and the key elements of the proposed work and reference papers. It then utilizes the reflective evaluator to filter the generated results. This process is interatively applied to each reference paper in the reference papers set until all reference papers are processed. The best result is ultimately retained as the model's generated output.

\subsubsection{Comparative Summarizer}
\label{sec:2.3.1}
For turn $i$, based on the proposed work key element $pro$,  the key element of the $i$-th reference paper $ref_i$  and comparative summarization guidance sequentially generated summary for each summary $s \in S_{i-1}$, and generating $n_s$ samples each time. $S_i = \left\{G(D_g,pro,ref_i,s,n_s),\forall s\in S_{i-1}\right\}$
Here, to enhance the comparability and organization of the generated summaries, comparative summarization guidance  are provided: \textit{"Considering the relationship between the reference paper and the target paper, as well as existing references in the previously completed related work, while retaining the content of all referenced papers mentioned in the previously completed related work."}

\subsubsection{Reflective Mechanism}
\label{sec:2.3.2}
Due to significant uncertainty in text generation tasks, we employ reflective generation to enhance the quality and stability of generated paragraphs. Here, we use LLMs as Reflective Evaluator to vote $n_v$ times on the generated results in each turn and then perform a statistical analysis on the voting results to obtain voting scores $E_i = E(D_e,S'_i)$.

Then we sort the scores, and retain the top $n_c$ candidates $S_i = \left\{S_t, t \in Sort(E_i)(1,n_c)\right\}$ . These selected candidates will be used for the next round of incremental generation. This approach helps identify the most promising results, ensures the quality of the generated text, and enhances generation stability.

\subsubsection{Reflective Incremental Generator Algorithm}
\label{sec:2.3.3}
In implementing reflective incremental generation, we drew inspiration from the breadth-first search algorithm for trees (Algorithm \ref{alg3}).
\begin{algorithm}[h]
    \caption{Reflective Incremental Generator} 
    \label{alg3} 
    \begin{algorithmic}
        \REQUIRE Proposed work key element $pro$, reference paper summaries list $refs\_list = \left [ref_1,ref_2,...ref_n\right]$, Comparative Summarizer $G()$, Reflective Evaluator $E()$, LM decoder for summarization $D_s$, LM decoder for evaluation $D_e$, n\_samples for each generation $n_s$, and the number of candidates retained for each turn is $n_c$.
        \STATE $S_0 \gets \left\{\right\}$ 
        \STATE $steps \gets len(refs\_list)$ 
        \FOR{$i = 1$ \TO $ steps $}
            \STATE $S'_t \gets \left\{G(D_g,pro,ref_i,s,n_s), s\in S_{i-1}\right\}$ 
            \STATE $E_i \gets E(D_e,S'_i)$ 
            \STATE $S_i \gets \left\{S_t, t \in Sort(E_i)(1,n_c)\right\}$ 
        \ENDFOR
        \RETURN $S_{argmax_iE_n(i)}$
    \end{algorithmic} 
\end{algorithm}

\noindent \textbf{notes:} $G()$ corresponds to the Comparative Summarizer function described in \S \ref{sec:2.3.1}, and $E()$ corresponds to the Reflective Envaluation function described in \S \ref{sec:2.3.2}. At each step, a collection containing $n_c$ most promising generated results is maintained, where the depth of the tree equals the number of documents in the relevant literature collection, $S'_t$ contains $n_c$ * $n_s$ results, while $S_{i-1}$ and $S_i$ each contain $n_c$ results. 

\section{G-Score: LLM-based automatic Evaluation Metrics}
The evaluation of generative tasks has always been challenging. Previous research on literature summarization predominantly depended on text summarization metrics, like ROUGE (\citet{Lin2004ROUGEAP}). However, conventional text summary evaluation metrics such as ROUGE fall short in gauging the quality of literature summaries. It is crucial to adopt more comprehensive evaluation criteria across various dimensions to guarantee that the generated literature summaries align with the necessary standards.
Here, inspired by G-Eval \cite{liu2023geval}, we attempted to assess it using LLMs. 
We established six-dimensional metrics for automatic evaluation based on research on literature summaries \cite{justitia2022automatic}. 

 \textbf{Evaluation Steps.} We used Large Language Models (LLMs) to score the six dimensions of generic quality and voted for the best summary from a series of model-generated summaries. Specially, to ensure fairness and consistency in evaluation, we simultaneously scored and voted for the generated results of multiple models in a single conversation. 

 \textbf{Evaluation Criterion:}

\noindent\textbf{Consistency (1-5)}: \textit{Content consistency between the generated summary and the gold summary. The generated summary must not contain content that conflicts with the gold summary.}

\noindent\textbf{Coherence(1-5)}: \textit{The quality of language coherence in generated summaries, which should not just be a heap of related information. }

\noindent\textbf{Comparative (1-5)}: \textit{Assess the extent to whether the generated summary conducts a comparative analysis on references and proposed work. Whether it provides an integrated summary of similar related works.}

\noindent\textbf{Integrity (1-5)}: \textit{Assess if the summary covers essential elements: research context, reference paper summaries, past research evaluation, contributions, and innovations.}

\noindent\textbf{Fluency (1-5)}: \textit{Assess the quality of the summary in terms of grammar, spelling, punctuation, word choice, and sentence structure.}

\noindent\textbf{Cite Accuracy(1-5)}: \textit{Assess whether the summary correctly cites reference paper in the format `[Reference i]' when mention the reference paper.}

\section{Experiment}
We validate the capabilities of our proposed \emph{ChatCite} agent by verifying the following questions: 1) Is the literature summary generated by \emph{ChatCite} better than that generated directly by LLMs with CoT and other LLM-based literature review approach? 2) Do all the modules in the \emph{ChatCite} contribute to its effectiveness? 3) What specific impact do the modules in the \emph{ChatCite} framework have on the quality of generated summary?

In this section, we conducted a series of experiments to address these questions. Firstly, we introduced our experimental setup (\S \ref{sec:4.1}). We compared the performance of existing large language models (LLMs) in directly generating related work under zero-shot and few-shot settings, as well as the best-performing LLM-based literature review approach (\S \ref{sec:4.2}). Additionally, we performed ablation analysis on each module in our agent to verify their respective capabilities (\S \ref{sec:4.3}). Finally, we conducted a human study for a detailed quality assessment of the generated related work summaries (\S \ref{sec:4.4}).
\subsection{Experimental Setup}
\label{sec:4.1}
\begin{table*}[t]
\centering
\small
  \renewcommand\arraystretch{1}
  \setlength{\tabcolsep}{4.0mm}{
  {
\begin{tabular}{l|ccc|c|c}
\toprule

\multirow{2}{*}{\centering Model}
& \multicolumn{3}{c|}{\textbf{R{\scriptsize OUGE} Metrics}} & \textbf{G-Score} & \textbf{G-Prf.} \\
\cmidrule(r){2-4}
& \multicolumn{1}{c|}{\textbf{R{\scriptsize OUGE}-1}}       
& \multicolumn{1}{c|}{\textbf{R{\scriptsize OUGE}-2}}       
& \multicolumn{1}{c|}{\textbf{R{\scriptsize OUGE}-L}}
& \multicolumn{1}{c|}{(1-5)}
& (\%)\\                        
\midrule
GPT-3.5 w/zero shot  & 26.01 & 6.11 & 24.02 & 3.4102 & 2.21 \\
GPT-3.5 w/few shot & 25.84 & 6.01 & 23.55 & 3.5968 & 10.80  \\
\midrule
GPT-4 w/zero shot & \textbf{30.02} &	\textbf{8.03} & \textbf{27.97} & 3.5076 &	26.40 \\
GPT-4 w/few shot & 15.52 & 1.78  & 14.20 & 1.6621 & 0.21 \\
LitLLM w/GPT-4 & 27.08 & 6.07 & 24.94 & 3.5448 &	24.51 \\
\midrule
 \textbf{ChatCite}  & 25.30  & 6.36  & 23.13 & \textbf{4.0642} & \textbf{35.86} \\
\midrule                 
\end{tabular}
}
\caption{\textbf{Main Results}: The results are automatically evaluated using ROUGE-1/2/L (F1) and the GPT-4.0 evaluator. G-Score represents the total score assessed by the GPT-4.0 evaluator, while G-Prf. indicates the model preferences among the five models.}
\label{tab:main_results}%
}
\end{table*}
\textbf{Dataset. }We conducted experiments to validate on a paper dataset NudtRwG-Citation dataset \cite{Wang2020ToCRWGET} designed for related work summarization task.
This test set includes 50 academic research papers in the field of Computer Science, each data containing the following components: 1) A target paper requiring related work generation without the related work section. 2) A ground truth related work section. 3) Reference papers of the target paper (annotated with authors and years). \par Each paper is well-received in conferences of computational linguistics and natural language processing, with an average citation number reaching 63.59, which indicates these target papers are widely recognized by the academic community.

\noindent\textbf{Models. }For the LLMs baseline, we employed the GPT-3.5 model (\citet{ouyang2022training}) with a 16k context window (version gpt-3.5-turbo-1106) and the GPT-4.0 model (\citet{achiam2023gpt}) with a 128K context window (gpt-4-turbo-preview). We evaluated their performance under zero-shot and few-shot settings.
For the previously best-performing LLM-based literature review approach, we use the recently proposed approach LitLLM \cite{agarwal2024litllm} as the baseline. We reproduce their ability to generate literature summaries according to the CoT prompt mentioned in their paper. To showcase its best performance, we use GPT-4.0 as the decoder for the LitLLM baseline.
For our model, due to the high cost of GPT-4.0, we conducted experiment based on GPT-3.5 (version gpt-3.5-turbo-1106) as the decoder for the experiment.
For evaluation, we use GPT-4.0 (gpt-4-turbo-preview) as decoder.

\noindent\textbf{Implementation. }In zero-shot setting, for GPT-3.5 model, due to the limitation of the context window, a two-step approach is used for generation: 1) summarizing and then generating with the prompt $\left[p_s\right]=$\textit{"Summarize the current article, preserving as much information as possible. Content:\{content\}"} for summarization. For generating the related work section, we use the prompt $\left[p_g\right]=$ \textit{"Generate the related work section based on the given target paper summary and its references summary. Read the Target Paper Content: \{Target\}. References content: \{References\}"}. For GPT-4.0 and LitLLM with GPT-4.0, $\left[p_g\right]$ is directly used for summarization.

In the few-shot setting, we add the instruction \textit{"Follow the writing style of the example but without including any content from the example. \{Examples\}"} to the zero-shot prompt.

\noindent\textbf{Evaluation metrics. }
We utilize both automatic metrics and human evaluations to assess the generic result. We employed traditional automatic metrics for summarization evaluation - the vocabulary overlap measures \textbf{ROUGE-1/2/L} (F1) (\citet{lin-2004-rouge}), our proposed LLM-based evaluation metrics G-Eval, and human evaluation under the same evaluation criterion.

\subsection{Main Results}
\label{sec:4.2}
We compared the performance of different baseline models on the paper test set (see Table \ref{tab:main_results}). In traditional summary evaluation metrics, such as ROUGE, GPT-4.0 achieved the best results under zero-shot settings. Although  ROUGE scores of \emph{ChatCite} may be slightly lower than GPT-4.0 with zero-shot, its performance in quality metrics generated by LLMs and the preference of LLMs is far superior to results obtained directly from other LLM baselines.

Surprisingly, GPT-4.0 performed poorly in few-shot settings.It is found that influenced by examples in the few-shot, resulting in irrelevant and erroneous summaries after case study. Notably, LitLLM with GPT-4.0 produced outcomes similar to GPT-4.0 in zero-shot but significantly lower than \emph{ChatCite}.

Therefore, we conclude that \textit{"ChatCite performs best among LLM-based literature summarization methods, and the approach following the human workflow guidance is superior to the results obtained by the Chain of Thought (CoT) method."}

\subsection{Ablation Analysis}
\label{sec:4.3}
Our proposed framework can be decomposed into two components: the Key Element Extractor and the Reflective Incremental Generator. The Reflective Incremental Generator comprises two key points: the Comparative Incremental Generation and the Reflective Mechanism. Therefore, we will analyze the three part separately.\par
\begin{table*}[t]
\centering
\small
  {
\begin{tabular}{l|ccc|c|c}
\toprule

\multirow{2}{*}{\centering Model}
& \multicolumn{3}{c|}{\textbf{R{\scriptsize OUGE} Metrics}} & \textbf{G-Score} & \textbf{G-Prf.} \\
\cmidrule(r){2-4}
& \multicolumn{1}{c|}{\textbf{R{\scriptsize OUGE}-1}}       
& \multicolumn{1}{c|}{\textbf{R{\scriptsize OUGE}-2}}       
& \multicolumn{1}{c|}{\textbf{R{\scriptsize OUGE}-L}}
& \multicolumn{1}{c|}{(1-5)}
& (\%)\\                        
\midrule
GPT-3.5 w/few shot  & \textbf{25.84} & 6.01 & \textbf{23.55} & 3.2426 & 2.84 \\
 -w/o Elem. & 24.38 & 5.81 & 22.36  & 4.0016 & 22.11\\
ChatCite -w/o Incre. & 24.72 & 5.93 & 22.40  & 3.8195 &	35.34 \\
\rowcolor{gray!20} \textbf{ChatCite}  & 25.30  & \textbf{6.36}  & 23.13 & \textbf{4.1064} & \textbf{39.71} \\
\midrule                 
\end{tabular}
}
\caption{\textbf{Ablation Results}: This table presents the ablation results on the model's Key Element Extractor and Comparative Incremental Generator, with the results of GPT-3.5 w/few-shot used as the baseline for GPT-3.5.}
\label{tab:ab_results}%

\end{table*}

\noindent \textbf{Key Element Extractor. }
To validate the effectiveness of the Key Element Extractor, we chose ChatCite without the Key Element Extractor as a comparison. The ChatCite without Key Element Extractor used the baseline summary prompt $\left[p_s\right]$ to directly summarize the article and then use Reflective Incremental Generator generate the literature summary.

In Table \ref{tab:ab_results}, comparing the results of ChatCite without Key Element Extractor and ChatCite, we can observe that ChatCite performs better in all dimensions of ROUGE metrics and the metrics generated by the LLM based evaluator. Therefore, it indicates that the Topic Extractor module plays an effective role in literature summarization.

\noindent \textbf{Comparative Incremental Mechanism.}
To validate the effectiveness of the Comparative Incremental Mechanism, we choose ChatCite without Comparative Incremental Mechanism as comparison, following the few-shot baseline prompt $\left[p_s\right]$ and few-shot examples as prompts to directly generate literature summaries from the text after standard summarization. Considering controlling variables for the incremental mechanism, we also incorporated CoT writing instructions into the method to ensure that the experimental results are not influenced by the writing instructions.

In Table \ref{tab:ab_results}, when comparing ChatCite with and without the Comparative Incremental Mechanism, the results indicate that ChatCite achieves higher ROUGE metrics and LLM-based evaluation metrics compared to ChatCite without the Comparative Incremental Mechanism. This suggests that the Comparative Incremental Mechanism significantly contributes to the effectiveness of literature summarization in the ChatCite framework.
\begin{figure}[ht]
    \includegraphics[width=1\linewidth]{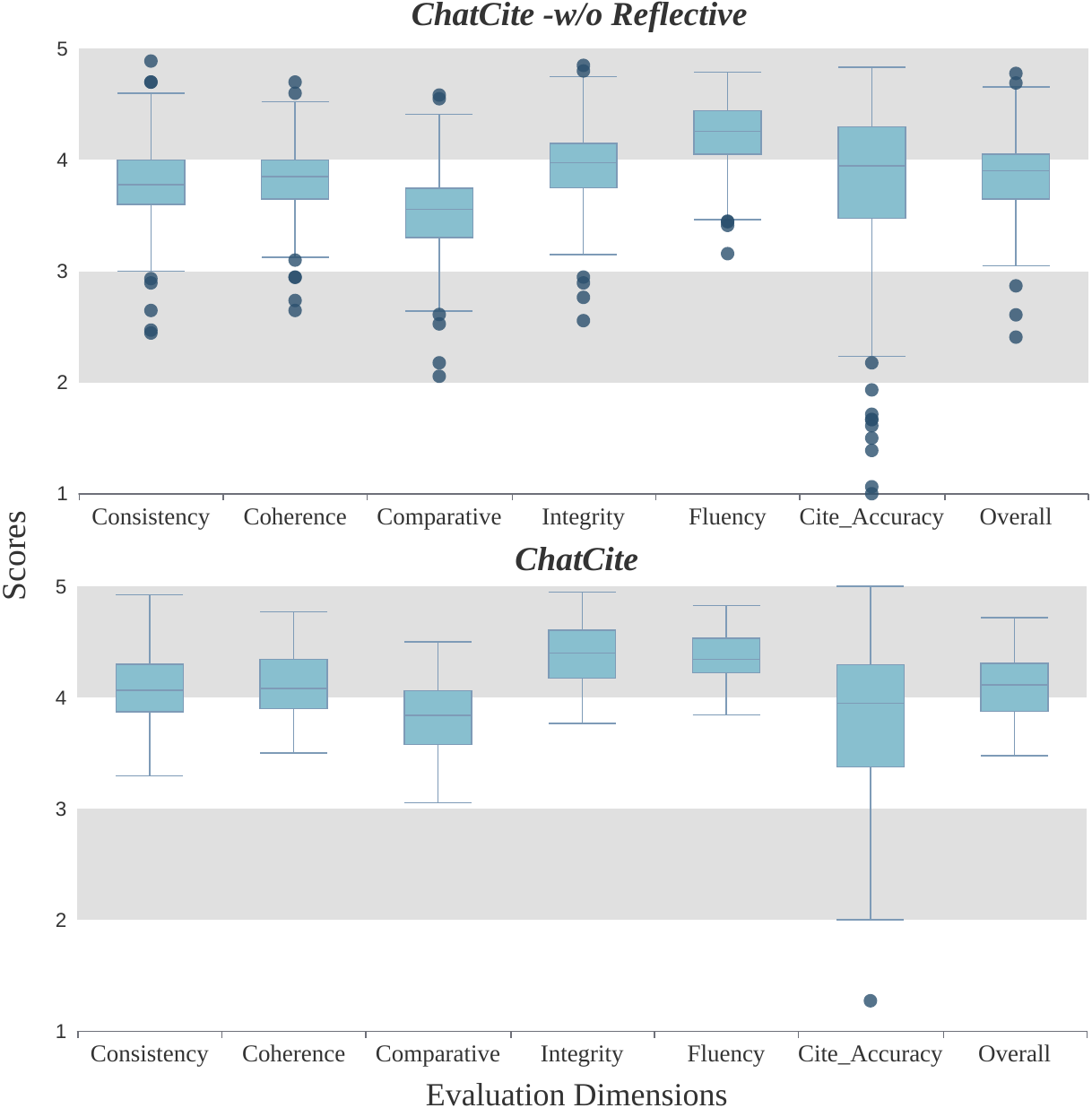}
    \caption{Ablation Study on the Reflective Mechanism. The upper and lower whiskers represent the overall range of the data, while the box displays the distribution of the middle 50\% of the dataset, with a line inside the box representing the median of the data. Data points outside the boxplot are considered outliers, indicating data points that significantly deviate from the box and whiskers. It can be observed that ChatCite performs more stable across all dimensions.}
    \label{fig:enter-label}
    \label{fig:rf_study}
\end{figure}

\noindent \textbf{Reflective Mechanism.}
In conclusion, we analyzed the reflective mechanism's impact. G-Scores for various dimensions were assessed based on multiple results from \emph{ChatCite}, both with and without the Reflective Mechanism. The boxplot results in Figure \ref{fig:rf_study} show similarities between the outcome of \emph{ChatCite} with and without the Reflective Mechanism. However, the overall results of \emph{ChatCite} are slightly higher, with minimal distribution outliers, suggesting a more stable generation of results. This affirms that the Reflective Mechanism effectively improves the quality and stability of the text generated in \emph{ChatCite}.

Overall, through ablation experiments on three components, we have demonstrated that \textit{"each part of \emph{ChatCite} framework  contributes to the improvement of the quality and stability of the generated results in literature summaries"}.
\subsection{Human Study}
\label{sec:4.4}
To conduct a fine-grained analysis on the quality of summary generated by \emph{ChatCite} and to understand the specific impact of individual modules on summarization, we conducted a human study. Several researchers in the field of computer science, with experience in academic writing, were enlisted to evaluate 10 selected samples using the same set of criteria and choose the better summary. 
\begin{figure}[h]
    \centering
    \includegraphics[width=1\linewidth]{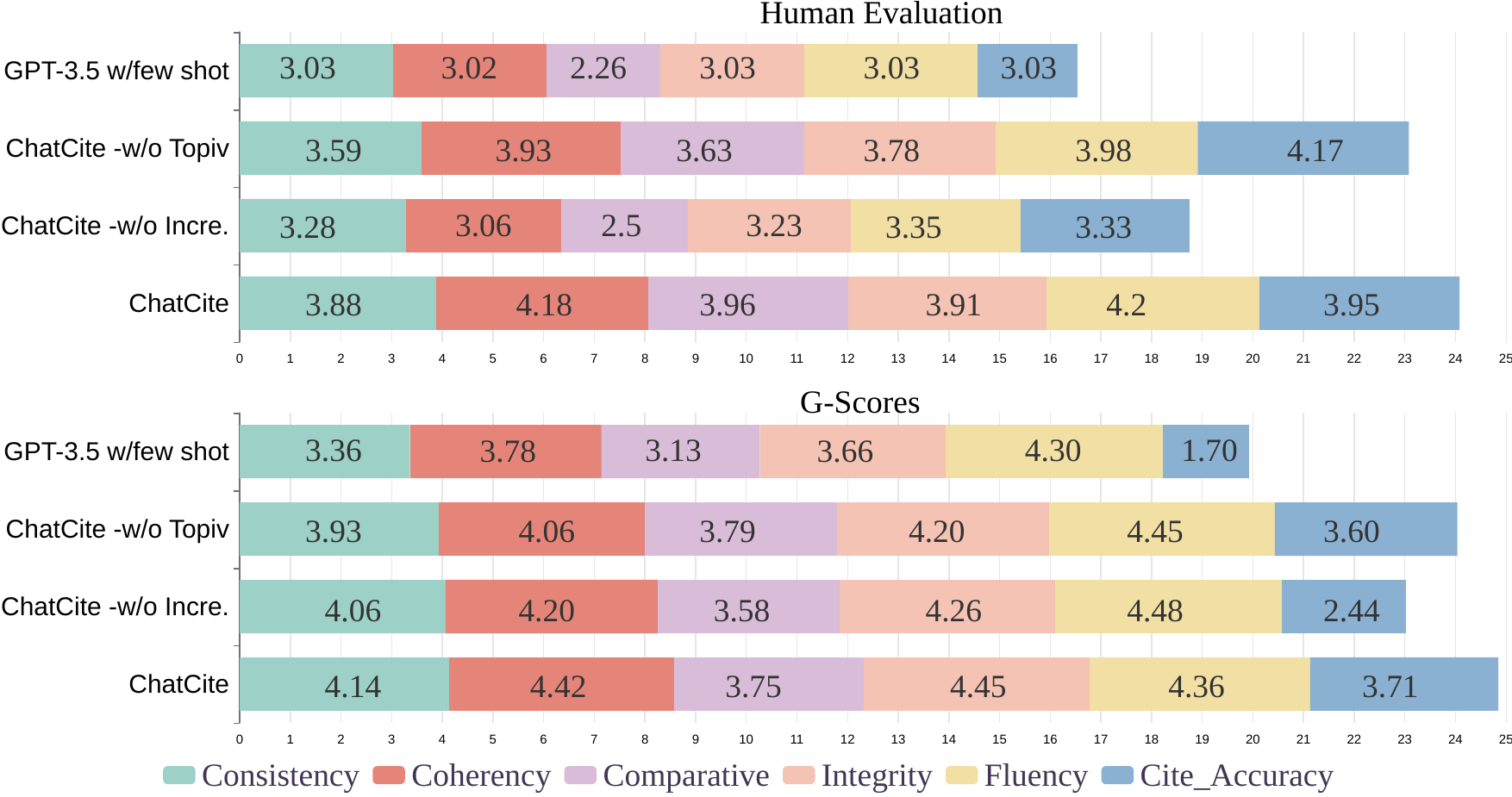}
    \caption{Human Evaluation \textit{vs.} G-Score on six dimensions of the generic summary quality. The scoring results of the G-Score model is aligned with the distribution of human evaluations.}
    \label{fig:4}
\end{figure}
\begin{figure}[h]
    \centering
    \includegraphics[width=1\linewidth]{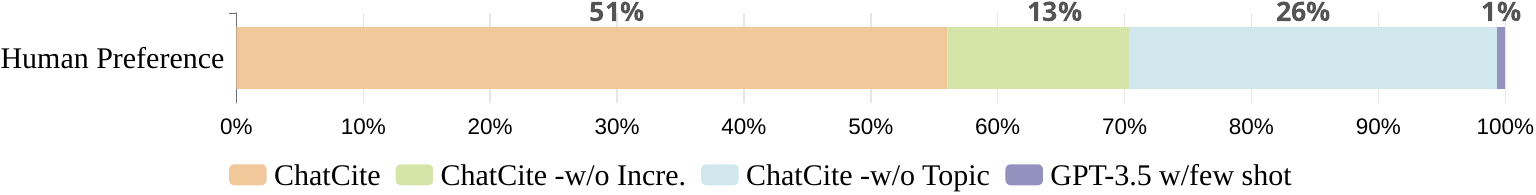}
    \caption{Human Preference: Average annotator vote distribution for better generated summaries.}
    \label{fig:prf}
\end{figure}

Figure \ref{fig:4} demonstrates the results of G-score metric align with human preferences. Specifically, the method incorporating Key Element Extractor exhibits higher content consistency. Summaries generated with the Comparative Incremental generation Mechanism demonstrate better characteristics of literature review, excelling in organizational structure, comparative analysis, and citation accuracy. The fluency of results generated by LLMs is consistently high, with relatively low variation among different models. In terms of human evaluation, summaries generated without the Comparative Incremental Mechanism exhibit overly discrete descriptions for each paper, lacking coherence. Unexpectedly, this feature was not captured in the assessment by the large models.

Additionally, Figure \ref{fig:prf} shows the extinct human preference of the ChatCite model over the others.
\section{Conclusion}
 LLMs are powerful tools in generating literature summaries, however, it poses the challenges of information omission, lack of comparative summaries and organizational deficiencies. In ChatCite, the Key Element Extractor contributes to improving content consistency, and the Comparative Incremental Generator effectively enhances the organizational structure, comparative analysis, and citation accuracy of the generated summary. Additionally, the literature summaries generated by ChatCite can be directly used for drafting literature reviews. Our study also demonstrated that the approach following the human workflow guidance is superior to the results obtained by the Chain of Thought (CoT) method. In the future, we hope that our work will further inspire research on complex inferential writing, enabling the full potential of LLMs in open-ended writing tasks.
\section*{Limitations}
In this work, we focused mainly on the summarization of specific topics based on the selected literatures instead of the collection and the filtering of the literatures themselves. The datasets primarily consist of research articles in the area of computer science and lack research articles from other fields of study to validate our model. Our experimentation used Chat GPT 3.5 as the tool for validating the quality of the generated content and the functionalities of the various components of the agent. We did not explore any additional spec that can influence the result of the GPT3.5 model nor the possibility of using other models as the validation tool. The evaluation of the generated content poses a great challenge. We evaluated the generated results from multiple dimensions using G-Score as the performance metric, but there is still room for improvements over the accuracy of the automatic evaluation process. The generated results exhibit randomness and instability. While our proposed approach demonstrates the effectiveness of the agent, the results have shown further research potential on improving the stability and quality of the output.

\section*{Ethics Statement}

The dataset we used consists of research articles sourced only from publicly available papers, eliminating concerns about data origin. We employ large language models as generators used and only used for summarizing people's ideas and literature and never on the innovative writing processes of the academic papers. However, if generated literature summaries are to be incorporated into academic paper writing, a review and editing of the generated results should be conducted. This ensures that academic writing content is free from harmful information and plagiarism issues.

We will make our code publicly available to ensure experiment reproducibility.

\bibliography{custom}
\bibliographystyle{acl_natbib}

\newpage
\appendix
\onecolumn

\section{Appendix}
\label{sec:appendix}
\subsection{An Example of generated results of all the models mentioned}
\begin{table*}[h]
\caption{An Example of literature summary results generated for Paper: \textit{BEL: Bagging for Entity Linking}}
\label{table:ex_gold}
\centering
\resizebox{\textwidth}{!}{ 
\scriptsize{ 
 \begin{tabular}{p{0.8\textwidth}}
  \hline
  \textbf{Gold literature Summary}\\\hline 		 
 Statistical machine translation systems often rely on large-scale parallel and monolingual training corpora to generate translations of high quality. Unfortunately, statistical machine translation system often suffers from data sparsity problem due to the fact that phrase tables are extracted from the limited bilingual corpus. Much work has been done to address the data sparsity problem such as the pivot language approach (Wu and Wang,2007; Cohn and Lapata, 2007) and deep learning techniques (Devlin et al., 2014; Gao et al., 2014; Sundermeyer et al., 2014; Liu et al., 2014).

 On the problem of how to translate one source language to many target languages within one model, few work has been done in statistical machine translation. A related work in SMT is the pivot language approach for statistical machine translation which uses a commonly used language as a ”bridge” to generate source-target translation for language pair with few training corpus. Pivot based statistical machine translation is crucial in machine translation for resource-poor language pairs, such as Spanish to Chinese. Considering the problem of translating one source language to many target languages, pivot based SMT approaches does work well given a large-scale source language to pivot language bilingual corpus and large-scale pivot language to target languages corpus. However, in reality, language pairs between English and many other target languages may not be large enough, and pivot-based SMT sometimes fails to handle this problem. Our approach handles one to many target language translation in a different way that we directly learn an end to multi-end translation system that does not need a pivot language based on the idea of neural machine translation.

 Neural Machine translation is a emerging new ﬁeld in machine translation, proposed by several work recently (Kalchbrenner and Blunsom, 2013; Sutskever et al., 2014; Bahdanau et al., 2014), aiming at end-to-end machine translation without phrase table extraction and language model training. Different from traditional statistical machine translation, neural machine translation encodes a variable-length source sentence with a recurrent neural network into a ﬁxed-length vector representation and decodes it with another recurrent neural network from a ﬁxed-length vector into variable-length target sentence. A typical model is the RNN encoder-decoder approach proposed by Bahdanau et al. (2014), which utilizes a bidirectional recurrent neural network to compress the source sentence information and ﬁts the conditional probability of words in target languages with a recurrent manner. Moreover, soft alignment parameters are considered in this model. As a speciﬁc example model in this paper, we adopt a RNN encoder-decoder neural machine translation model for multi-task learning, though all neural network based model can be adapted in our framework.

 In the natural language processing ﬁeld, a1724 notable work related with multi-task learning was proposed by Collobert et al. (2011) which shared common representation for input words and solve different traditional NLP tasks such as part-of-Speech tagging, name entity recognition and semantic role labeling within one framework, where the convolutional neural network model was used. Hatori et al. (2012) proposed to jointly train word segmentation, POS tagging and dependency parsing, which can also be seen as a multi-task learning approach. Similar idea has also been proposed by Li et al. (2014) in Chinese dependency parsing. Most of multi-task learning or joint training frameworks can be summarized as parameter sharing approaches proposed by Ando and Zhang (2005) where they jointly trained models and shared center parameters in NLP tasks. Researchers have also explored similar approaches (Sennrich et al., 2013; Cui et al., 2013) in statistical machine translation which are often refered as domain adaption. Our work explores the possibility of machine translation under the multitask framework by using the recurrent neural networks. To the best of our knowledge, this is theﬁrst trial of end to end machine translation under multi-task learning framework.
\\ \hline
\end{tabular}
}}
\end{table*}
\begin{table*}[h]
\label{table:ex}
\centering
\resizebox{\textwidth}{!}{ 
\scriptsize{ 
 \begin{tabular}{p{0.8\textwidth}}
  \hline
 \textbf{ChatCite with GPT-3.5}\\\hline 		
 Related Work

The field of machine translation has witnessed remarkable progress in addressing data scarcity and enhancing translation quality. Early approaches such as the pivot language approach proposed by Wu and Wang (2007) [Reference 1], and the triangulation method introduced by Cohn and Lapata (2007) [Reference 2], focused on improving phrase-based statistical machine translation, particularly for language pairs with scarce resources. Additionally, Devlin et al. (2014) [Reference 3] contributed by introducing a neural network joint model (NNJM) for machine translation, offering valuable insights into the formulation of a novel neural network joint model and techniques for enhancing machine translation performance.

Moreover, Gao et al. (2014) [Reference 4] enriched the understanding of addressing data scarcity and improving translation quality by introducing a Continuous-space Phrase Translation Model (CPTM) to project source and target phrases into continuous-valued vector representations, complementing the approaches of previous researchers. Furthermore, Sundermeyer et al. (2014) [Reference 5] expanded the scope of machine translation by investigating the effectiveness of recurrent neural networks in translation modeling, contributing to the broader landscape of machine translation research.

In the context of the target paper, which proposed a multi-task learning framework for machine translation, the works by Zhenghua Li et al. (2011) [Reference 6] and Nal Kalchbrenner and Phil Blunsom (2013) [Reference 7] proved highly relevant. Zhenghua Li et al. (2011) proposed joint optimization for Chinese POS tagging and dependency parsing, offering valuable insights into joint optimization techniques. Similarly, Nal Kalchbrenner and Phil Blunsom (2013) [Reference 7] introduced recurrent continuous translation models (RCTMs), providing a new perspective in the domain of machine translation and contributing to the advancement of purely continuous sentence-level translation models.

The work by Ilya Sutskever, Oriol Vinyals, and Quoc V. Le (2014) [Reference 8] on sequence to sequence learning with neural networks is also highly relevant to the target paper's objectives. The paper addresses the challenge of mapping sequences to sequences using Deep Neural Networks (DNNs) and proposes a general end-to-end approach to sequence learning that makes minimal assumptions on the sequence structure, aligning with the multi-task learning framework proposed in the target paper.

Additionally, the 9th reference paper by Ronan Collobert et al. (2011) [Reference 9] addresses the effectiveness of leveraging large unlabeled datasets and multitask supervised training to improve performance, aligning with the multi-task learning framework proposed in the target paper. The recently explored Incremental Joint Approach to Word Segmentation, POS Tagging, and Dependency Parsing in Chinese by Jun Hatori, Takuya Matsuzaki, Yusuke Miyao, and Jun (2012) [Reference 10] presents a joint model for word segmentation, POS tagging, and dependency parsing for Chinese, demonstrating significant improvements in accuracies of the three tasks over baseline models. The study's positive results in Chinese segmentation accuracies and the effectiveness of the proposed joint model are highly relevant to the advancements in multi-task learning frameworks for machine translation, providing further insights into joint model approaches and their potential for improving translation quality.

The 11th reference paper by Rie Kubota Ando and Tong Zhang (2005) focuses on learning predictive structures from multiple tasks and unlabeled data. It attempts to improve the performance of supervised learning algorithms by including unlabeled data and investigates a closely related problem that leads to a novel approach to semi-supervised learning. The paper introduces a general framework for structural learning and proposes algorithms for structural learning, demonstrating the effectiveness of the proposed algorithms in semi-supervised learning.

In a similar vein, the proposed multi-task learning framework in the target paper aims to improve the translation quality and address data scarcity by leveraging multiple tasks and large-scale parallel training corpora. By integrating insights from Rie Kubota Ando and Tong Zhang (2005) [Reference 11] with existing literature, the target paper aligns with the broader trend of leveraging multiple tasks and unlabeled data to enhance machine learning models. The contributions of both the target paper and the referenced works collectively enrich the understanding of addressing data scarcity, enhancing translation quality, and advancing the capabilities of machine translation models, laying the foundation for the multi-task learning framework proposed in the target paper. These contributions provide valuable insights into the advancements in machine translation, emphasizing contributions to the research field and predicting potential future directions.
\\ \hline
  \end{tabular}
}}
\end{table*}
\begin{table*}[h]

\label{table:this}
\centering
\resizebox{\textwidth}{!}{ 
\scriptsize{ 
 \begin{tabular}{p{0.8\textwidth}}
  \hline
\textbf{ChatCite without Key Elements Extractor}\\\hline
Related Work

The field of machine translation has witnessed significant advancements in overcoming challenges related to limited training resources and enhancing translation quality across various language pairs. Notably, approaches such as the pivot language method, multi-parallel corpora, and the triangulation method have been explored to address scenarios with limited training data [Radford et al. (2019)[Reference 5]]. Additionally, the development of neural network joint models (NNJM) [Sutskever, Vinyals, and Le (2014)[Reference 8]] and the integration of bidirectional recurrent neural networks have shown promise in improving translation quality, particularly for language pairs with limited resources.

In alignment with these developments, the target paper focuses on the translation of sentences from a source language to multiple target languages using a multi-task learning framework inspired by neural machine translation. This approach not only demonstrates substantial progress in machine translation, especially for languages with limited training resources but also introduces a neural network joint model (NNJM) for statistical machine translation, aligning with the innovative approaches presented in the referenced studies and offering promise for enhancing translation quality for language pairs with limited resources.

Moreover, \"Incremental joint approach to word segmentation, POS tagging, and dependency parsing in Chinese\" [Reference 10] proposes a joint model for word segmentation, POS tagging, and dependency parsing in Chinese, showing improvements particularly for out-of-vocabulary words. This study contributes to a comprehensive understanding of the methods used to address challenges in machine translation and natural language processing.

Expanding on these foundations, the 11-th reference paper introduces a framework for learning predictive structures from multiple tasks and unlabeled data. This framework presents a general approach for formulating and analyzing the structural learning problem in the context of learning with unlabeled data, demonstrating its relevance to the challenges in machine translation. The proposed algorithm for joint empirical risk minimization is particularly notable, as it addresses the problem of learning predictive structures on hypothesis spaces from multiple learning tasks, aligning with the multi-task learning framework proposed in the target paper. Furthermore, the experiments presented in the 11-th reference show significant performance improvements over supervised methods in various tasks, indicating the effectiveness of the proposed framework in semi-supervised learning settings.

In summary, the integrated literature on machine translation and natural language processing contributes valuable insights for the development of machine translation systems, offering innovative approaches to enhance translation quality, particularly for language pairs with limited resources. The introduction of the framework for learning predictive structures from multiple tasks and unlabeled data adds a new dimension to the existing research landscape, emphasizing the potential for semi-supervised learning to address challenges in machine translation and natural language processing. These advancements collectively pave the way for potential future research directions in the field of machine translation.
\\ \hline
  \end{tabular}
}}
\end{table*}
\begin{table*}[h]

\label{table:this}
\centering
\resizebox{\textwidth}{!}{ 
\scriptsize{ 
 \begin{tabular}{p{0.8\textwidth}}
  \hline
\textbf{ChatCite without Reflective Incremental Generator}\\\hline
Related Work

The related work section provides an overview of the literature relevant to the target paper's research topic, which aims to address the problem of learning a machine translation model that can simultaneously translate sentences from one source language to multiple target languages. The related work encompasses various studies in the field of machine translation, natural language processing, and neural network models.

[1] Hua Wu and Haifeng Wang (2007) proposed a pivot language approach for phrase-based statistical machine translation, which addresses the translation problem for language pairs with scarce resources by using a pivot language and making use of large bilingual corpora without language-dependent resources or deep linguistic processing. The study demonstrated the effectiveness of the pivot language method for translation on language pairs with a small bilingual corpus.

[2] Trevor Cohn and Mirella Lapata (2007) introduced the method of triangulation for translation modeling, which translates from a source to a target language via an intermediate third language, to exploit multi-parallel corpora for training and improve the coverage and quality of phrase-based statistical machine translation. The research focused on addressing the issue of poor performance of current phrase-based SMT systems when using small training sets.

[3] Jacob Devlin et al. (2014) formulated a neural network joint model (NNJM) for machine translation, along with techniques to overcome the high cost of using NNLM-style models in MT decoding. The study demonstrated significant improvements in machine translation performance using the proposed NNJM and its variations.

[4] Jianfeng Gao et al. (2014) introduced the Continuous-space Phrase Translation Model (CPTM) to address the sparsity problem in estimating phrase translation probabilities by learning continuous phrase representations. The study demonstrated substantial improvement over the baseline system with a statistically significant margin.

[5] Martin Sundermeyer et al. (2014) explored the effectiveness of recurrent neural networks in translation modeling, specifically focusing on word-based and phrase-based approaches, as well as the inclusion of bidirectional architectures. The research demonstrated improvements over strong baselines in translation modeling.

[6] Zhenghua Li et al. (2011) proposed a joint optimization approach for Chinese POS tagging and dependency parsing, showcasing significant improvements in parsing accuracy. The study addressed the issue of error propagation in parsing accuracy due to using automatic POS tags instead of gold ones.

[7] Nal Kalchbrenner and Phil Blunsom (2013) introduced Recurrent Continuous Translation Models (RCTMs), demonstrating their translation capabilities and low perplexities with respect to reference translations. The study highlighted the sensitivity of RCTMs to word order, syntax, and meaning despite lacking alignments.

[8] Ilya Sutskever et al. (2014) presented a general end-to-end approach to sequence learning using Deep Neural Networks (DNNs) for mapping sequences to sequences, showcasing the outperformance of the proposed LSTM-based approach over a mature SMT system.

[9] Ronan Collobert et al. (2011) proposed a unified neural network architecture and learning algorithm for various natural language processing tasks, demonstrating the effectiveness of leveraging large unlabeled datasets and multitask supervised training to improve performance.

[10] Jun Hatori et al. (2012) introduced an incremental joint approach to word segmentation, POS tagging, and dependency parsing for Chinese, showcasing substantial improvements in the accuracies of the three tasks over baseline models.

[11] Rie Kubota Ando and Tong Zhang (2005) proposed a framework for learning predictive structures from multiple tasks and unlabeled data, demonstrating significant performance improvements over supervised methods in various tasks through structural learning.

These studies collectively contribute to the field of machine translation, natural language processing, and neural network models, providing insights and advancements that are relevant to the target paper's research on learning a machine translation model for translating sentences from one source language to multiple target languages. The innovations and findings from these studies serve as valuable background and insights for the target research, highlighting the significance of the related work to the target study and potential future research directions in the field.
\\\hline
 \end{tabular}
}}
\end{table*}
\begin{table*}[h]
\label{table:this}
\centering
\resizebox{\textwidth}{!}{ 
\scriptsize{ 
 \begin{tabular}{p{0.8\textwidth}}
  \hline
\textbf{GPT-3.5 with zero shot}\\\hline
Related Work Section:

The proposed multi-task learning framework for translating sentences from one source language to multiple target languages using a machine translation model builds upon and is related to several key works in the field of machine translation and natural language processing.

1. Pivot language approach for phrase-based statistical machine translation

The article \"Pivot language approach for phrase-based statistical machine translation\" proposes a method for phrase-based statistical machine translation using a pivot language to address translation problems for language pairs with scarce resources. This method is related to the challenge of translating between languages with limited parallel data, which is also addressed in the target paper. The pivot language approach demonstrates the effectiveness of utilizing large bilingual corpora from a pivot language to improve translation quality, especially for language pairs with limited training data.

2. Machine translation by triangulation: Making effective use of multi-parallel corpora

The article discusses the method of triangulation, which involves translating from a source to a target language via an intermediate third language, aiming to improve machine translation systems by exploiting multi-parallel corpora. This work is relevant to the target paper as it addresses the issue of improving translation quality using a wider range of parallel corpora, particularly with small training sets and for translating between low-density language pairs.

3. Fast and robust neural network joint models for statistical machine translation

The article presents a novel neural network joint model (NNJM) for statistical machine translation, which is related to the use of neural network models in machine translation. The NNJM augments a neural network language model with a source context window and is purely lexicalized, demonstrating significant improvements in BLEU scores. This work is relevant to the target paper's exploration of neural machine translation and the effectiveness of neural network-based models in improving translation quality.

4. Learning continuous phrase representations for translation modeling

The article introduces a novel phrase translation model, the Continuous-space Phrase Translation Model (CPTM), which addresses the sparsity problem in estimating phrase translation probabilities. This work is related to the target paper's exploration of continuous representations and their use in improving translation accuracy. The CPTM significantly improves the accuracy of a state-of-the-art phrase-based SMT system, demonstrating its effectiveness in improving translation quality.

5. Translation modeling with bidirectional recurrent neural networks

The article presents two different translation models using recurrent neural networks, demonstrating their capability to improve machine translation performance. This work is related to the target paper in showcasing the effectiveness of recurrent neural networks in improving translation quality, particularly with the use of bidirectional recurrent models.

6. Joint optimization for Chinese POS tagging and dependency parsing

The article discusses the importance of joint optimization for Chinese POS tagging and dependency parsing, proposing a unique model to optimize both tasks. This work is related to the target paper's exploration of joint learning frameworks and their impact on improving parsing accuracy, demonstrating the potential of joint optimization for improving translation quality.

7. Recurrent continuous translation models

The article introduces a class of probabilistic continuous translation models called Recurrent Continuous Translation Models (RCTMs), which are purely based on continuous representations for words, phrases, and sentences. This work is related to the target paper's exploration of continuous representations and their potential in estimating translation quality, highlighting the effectiveness of RCTMs in capturing syntactic and semantic information.

8. Sequence to sequence learning with neural networks

The article discusses the use of Deep Neural Networks (DNNs) for sequence to sequence learning, particularly in the context of machine translation. This work is related to the target paper's exploration of neural network-based models and their effectiveness in improving translation performance, demonstrating the potential of sequence to sequence learning with neural networks for various sequence to sequence problems.

9. Incremental joint approach to word segmentation, POS tagging, and dependency parsing in Chinese

The article proposes a joint model for word segmentation, POS tagging, and dependency parsing in Chinese, demonstrating significant improvements in accuracies for all three tasks over baseline models. This work is related to the target paper's exploration of joint learning frameworks and their impact on improving translation quality, particularly in the context of joint modeling for multiple language translation tasks.

10. A framework for learning predictive structures from multiple tasks and unlabeled data

The article presents a framework for learning predictive structures from multiple tasks and unlabeled data, addressing the concept of semi-supervised learning and the problem of learning predictive structures on hypothesis spaces from multiple learning tasks. This work is related to the target paper's exploration of multi-task learning frameworks and their effectiveness in improving translation quality, particularly in the context of learning from multiple tasks and unlabeled data.

The related work presented here demonstrates the diverse approaches and techniques employed in the field of machine translation and natural language processing, providing valuable insights and methods that contribute to the advancement of multi-task learning frameworks and neural machine translation models for improving translation quality.
\\ \hline
  \end{tabular}
}}
\end{table*}
\begin{table*}[h]

\label{table:this}
\centering
\resizebox{\textwidth}{!}{ 
\scriptsize{ 
 \begin{tabular}{p{0.8\textwidth}}
  \hline
\textbf{GPT-3.5 with few shot}\\\hline
Related Work Section:

The related work section will include a review of existing literature and studies that are relevant to the topic of multi-task learning for machine translation and related methods for improving translation quality. The section will discuss the following papers:

1. \"Pivot language approach for phrase-based statistical machine translation\" - This study proposes a new method for phrase-based statistical machine translation using a pivot language to address translation problems for language pairs with scarce resources. The method demonstrates the use of large bilingual corpora from a pivot language to improve translation quality for language pairs with limited training data.

2. \"Machine translation by triangulation: Making effective use of multi-parallel corpora\" - The article discusses a method for improving machine translation systems by exploiting multi-parallel corpora. The proposed method, called triangulation, involves translating from a source to a target language via an intermediate third language, allowing the use of a wider range of parallel corpora for training.

3. \"Fast and robust neural network joint models for statistical machine translation\" - This article discusses the development of a novel neural network joint model (NNJM) for statistical machine translation (SMT). The NNJM augments a neural network language model (NNLM) with a source context window and demonstrates significant improvements in translation quality.

4. \"Learning continuous phrase representations for translation modeling\" - The study presents a novel phrase translation model for statistical machine translation (SMT) that addresses the sparsity problem in estimating phrase translation probabilities. The model utilizes continuous phrase representations to improve the accuracy of a state-of-the-art phrase-based SMT system.

5. \"Translation modeling with bidirectional recurrent neural networks\" - The article introduces two different translation models using recurrent neural networks and demonstrates their effectiveness in improving machine translation performance on various tasks.

6. \"Recurrent continuous translation models\" - This study introduces a class of probabilistic continuous translation models called Recurrent Continuous Translation Models (RCTMs) and demonstrates their potential in capturing syntactic and semantic information and estimating translation quality.

7. \"Sequence to sequence learning with neural networks\" - The article discusses the use of Deep Neural Networks (DNNs) for sequence to sequence learning and demonstrates the effectiveness of this approach on an English to French translation task.

8. \"Incremental joint approach to word segmentation, POS tagging, and dependency parsing in Chinese\" - The article proposes a joint model for word segmentation, POS tagging, and dependency parsing in Chinese and demonstrates significant improvements in accuracies for all three tasks over baseline models.

9. \"A framework for learning predictive structures from multiple tasks and unlabeled data\" - The study presents a framework for learning predictive structures from multiple tasks and unlabeled data, demonstrating significant performance improvements over supervised methods in various text categorization, named entity chunking, part-of-speech tagging, and hand-written digit image classification tasks.
\\\hline
 \end{tabular}
}}
\end{table*}
\begin{table*}[h]
\label{table:this}
\centering
\resizebox{\textwidth}{!}{ 
\scriptsize{ 
 \begin{tabular}{p{0.8\textwidth}}
  \hline
\textbf{GPT-4.0 with zero shot}\\\hline
Based on the given information, here is a related work section for the target paper:

Related Work

The problem of learning from multiple tasks and utilizing unlabeled data has been an area of significant research interest in machine learning and natural language processing (NLP). The concept of semi-supervised learning, which involves the use of both labeled and unlabeled data, has been explored in various studies with the aim of improving the performance of supervised learning algorithms. Vapnik introduced the notion of transductive inference, which is a form of semi-supervised learning, although its effectiveness has been debated (Vapnik, 1998; Zhang and Oles, 2000). Co-training, another semi-supervised approach, was proposed by Blum and Mitchell (1998) and is related to the bootstrap method used in NLP applications (Yarowsky, 1995). However, the effectiveness of co-training can be limited by the assumptions made by the method and the potential introduction of noise into the labels (Pierce and Cardie, 2001).

The idea of learning predictive structures from multiple tasks, also known as structural learning or multi-task learning, has been discussed in the literature. Ando and Zhang (2005) presented a framework for structural learning where multiple tasks are used to discover a shared low-dimensional predictive structure. This approach is based on the principle that a good hypothesis space should have a small approximation error and a small estimation error. The authors proposed an algorithm for structural learning that finds a common low-dimensional feature space shared by multiple problems. Their method was shown to be effective in semi-supervised learning settings, improving the performance of various NLP tasks.

In the context of NLP, the joint approach to tasks such as word segmentation, POS tagging, and dependency parsing has been explored. Hatori et al. (2012) proposed an incremental joint model for Chinese POS tagging and dependency parsing, demonstrating improvements over baseline models. The work by Kubota Ando and Zhang (2005) extends this idea to a joint model that simultaneously processes word segmentation, POS tagging, and dependency parsing, capturing global interactions among these tasks. Their model employs an efficient character-based decoding method that combines features from state-of-the-art segmentation, POS tagging, and dependency parsing models.

The use of external resources, such as dictionaries, has also been considered in joint models to enhance performance. Wang et al. (2011) used semi-supervised methods with large auto-analyzed data to improve Chinese word segmentation and POS tagging. The incorporation of syntactic information has been shown to improve the segmentation of out-of-vocabulary words, which is a common challenge in languages like Chinese where transliterated words often include characters used as common or function words.

Overall, the related work demonstrates the potential of joint and semi-supervised approaches in leveraging multiple tasks and unlabeled data to improve the performance of machine learning models in NLP tasks. The target paper builds upon these concepts by proposing a unified framework that avoids task-specific engineering and learns internal representations from vast amounts of unlabeled data, achieving good performance across various NLP tasks.
\\ \hline
 \end{tabular}
}}
\end{table*}

\begin{table*}[h]
\label{table:this}
\centering
\resizebox{\textwidth}{!}{ 
\scriptsize{ 
 \begin{tabular}{p{0.8\textwidth}}
  \hline
\textbf{GPT-4.0 with few shot}\\\hline
The examples provided discuss the use of Large Language Models (LLMs) as proxies for human behaviors and the systematic study of LLMs' personalities based on established personality trait theories. Here's a summary of the key points from the examples:

1. LLMs Mimicking Human Behaviors:

- Recent studies have shown that LLMs can mimic human behaviors, including reasoning and cognitive abilities, as well as participate in social science experiments.

- These studies are mostly empirical and based on case-by-case observations.

2. Systematic Study of LLMs' Personalities:

  - The authors propose a systematic and quantitative approach to study LLMs' behaviors by evaluating and inducing personality traits in LLMs.

  - This approach does not require supervised fine-tuning or human evaluation of generated utterances.

3. Personality Trait Theories:

  - The Big Five and Sixteen Personality Factors (16PF) are two prominent theories that provide consistent descriptions of individual differences.

  - Psychometric tests based on these theories, such as NEO-PI-R, are effective in assessing personality traits and have shown that human differences can be categorized into continuous factor dimensions.

4. Correlation Between Personality and Behavior:

  - Empirical studies have confirmed a strong correlation between personality traits and real-world human behaviors.
  
  - Specifically, there is a significant link between the Big Five traits and language use, suggesting that personality influences how people communicate.
  
The examples highlight the potential of LLMs to simulate complex human-like behaviors and the importance of understanding the personality traits of these models for various applications.
\\ \hline
 \end{tabular}
}}
\end{table*}

\begin{table*}[h]
\label{table:this}
\centering
\resizebox{\textwidth}{!}{ 
\scriptsize{ 
 \begin{tabular}{p{0.8\textwidth}}
  \hline
\textbf{LitLLM with GPT-4.0}\\\hline
The Unified Neural Network Model for NLP Tasks

Natural Language Processing (NLP) tasks such as part-of-speech (POS) tagging, chunking, named entity recognition (NER), and semantic role labeling (SRL) are essential for understanding and generating human language. Traditional approaches to these tasks often involve task-specific feature engineering, which can be labor-intensive and may not generalize well to new tasks or datasets.

To address these issues, researchers have proposed unified neural network models that can learn to perform multiple NLP tasks simultaneously without extensive feature engineering. These models leverage large amounts of both labeled and unlabeled data to learn useful representations for words, phrases, and sentences that capture syntactic and semantic information.

One such model is the Long Short-Term Memory (LSTM) network, which has been shown to be effective for sequence-to-sequence learning tasks like machine translation. LSTMs can process input sequences of variable lengths and learn long-range dependencies, making them suitable for NLP tasks. They can be trained end-to-end with backpropagation and can learn from both supervised and unsupervised data.

Another approach is the use of semi-supervised learning methods, which combine labeled and unlabeled data to improve model performance. For example, structural learning algorithms can learn predictive structures from multiple tasks and use them to enhance the performance of individual tasks. This can be particularly useful when there is a large amount of unlabeled data available.

The success of these unified neural network models demonstrates the potential of using deep learning techniques for NLP tasks. By avoiding task-specific engineering and leveraging large datasets, these models can achieve state-of-the-art performance on a variety of benchmarks, making them valuable tools for researchers and practitioners in the field of NLP."
\\ \hline 

  \end{tabular}
}}

\end{table*}

\clearpage 
\subsection{Related work draft for this paper generated by ChatCite with GPT-4.0}

\begin{table*}[h]
\caption{We show the related work summary using this paper and our collection of related work papers generated by the ChatCite model based on GPT-4.0 (gpt-4-turbo-preview version). It can be observed that the citations in the paper are consistent with those in the collection, the content is accurate at the same time. We utilized this summary as a draft and, completed our related work section after some minor modifications.}
\label{table:this_rw}
\centering
\resizebox{\textwidth}{!}{ 
\scriptsize{ 
 \begin{tabular}{p{0.8\textwidth}}
  \hline
  \textbf{Generated literature summary for this paper}\\\hline 		 
In the evolving landscape of automated scholarly writing, particularly in the domain of literature review summarization, the proposed ChatCite framework within our target paper represents a significant advancement, aiming to harness the capabilities of Large Language Models (LLMs) to generate high-quality related work summaries. The foundational work in this field began with hierarchical keyword structures for extractive summarization [Reference 1], and has progressed through various iterations including multi-source pointer-generator networks [Reference 2], and sophisticated models like the ARWG system [Reference 3], BACO [Reference 4], and the Relation-aware Related work Generator (RRG) [Reference 5]. The introduction of contrastive learning to improve summarization quality [Reference 6], and automatic citation sentence generation methods [Reference 7], have further refined the capabilities of LLMs in this space.

Recent studies have concentrated on the unsupervised and few-shot learning capabilities of LLMs, notably GPT-3, establishing a solid groundwork for AI application in scholarly communication [References 8-9]. The exploration of AI tools such as ChatGPT by Huang, Jingshan, and Ming Tan (2023) [Reference 10] in crafting scientific review articles, sheds light on both the potential benefits and drawbacks of AI in academic writing, such as enhancement in speed and quality versus risks of inaccuracies and bias. Building on these insights, the LitLLM toolkit suggested by Shubham Agarwal, Issam H. Laradji et al. (2024) [Reference 11] brings novel approaches to overcome challenges like the generation of hallucinated content and the overlooking of recent studies by employing Retrieval Augmented Generation (RAG) principles, specialized prompting, and instructive techniques. This showcases a significant alignment with our target paper's objectives, particularly in mitigating hallucinations and enhancing summarization accuracy through the ChatCite framework.

The latest contribution in this domain is presented by Yang Liu, Dan Iter, et al. (2023) [Reference 12] through the G-EVAL framework, which introduces a methodology for evaluating the quality of texts generated by natural language generation (NLG) systems using LLMs, leveraging chain-of-thoughts (CoT) and a form-filling paradigm. This approach, which significantly outperforms previous methods in evaluating the creativity and diversity required in text generation tasks, aligns closely with the goals of the ChatCite framework, especially in its pursuit of generating high-quality literature summaries. The G-EVAL framework not only surpasses all previous state-of-the-art evaluators on the SummEval and Topical-Chat benchmarks but also underscores the importance of instruction sensitivity and the benefits of a CoT approach, which could potentially inform future enhancements in the ChatCite framework [Reference 12].

By evaluating the quality of NLG outputs, [Reference 12] addresses the limitations identified in prior works, including the potential bias of LLM-based evaluators towards LLM-generated texts, and calls for further research to understand the behavior of LLM-based evaluators fully. This contribution is particularly relevant to our target paper, which aims to enhance the quality and stability of literature review summarizations generated by LLMs. Thus, integrating insights from G-EVAL [Reference 12] can provide a more nuanced understanding of evaluating the effectiveness of automated scholarly writing tools, highlighting a potential area for future development within the scope of the ChatCite framework.

In summary, the development of automated scholarly writing, particularly in literature review summarization, has progressed through key innovations in LLM capabilities, methodological enhancements, and now, through improved evaluation frameworks. This journey from extractive summarization to sophisticated LLM-based frameworks underscores a collaborative advancement in the field, with each contribution building upon the last to address emerging challenges and achieve greater accuracy, efficiency, and relevance in automated scholarly writing. The ChatCite framework, supported by insights from references [References 1-12], continues this trajectory by proposing innovative solutions to the challenges in literature review summarization, setting a platform for further exploration and innovation in the field.
\\ \hline 
  \end{tabular}
  }}
\end{table*}

\end{document}